\documentclass[preprint,showpacs,preprintnumbers,amsmath,amssymb]{revtex4}
\input psfig.sty

\usepackage{graphicx}
\usepackage{dcolumn}
\usepackage{bm}

\begin{document}


\title{Constraining the Nonextensive Mass Function of Halos from BAO, CMB and X-ray data}

\author{L. Marassi\footnote{luciomarassi@astro.iag.usp.br}\,, J. V. Cunha\footnote{cunhajv@astro.iag.usp.br}\,, J. A. S.
Lima\footnote{limajas@astro.iag.usp.br}}


\affiliation{Instituto de Astronomia, Geof\'{\i}sica e Ci\^encias
Atmosf\'ericas, USP, CEP 05508-090, S\~ao Paulo, SP, Brasil}

\date{\today}

\begin{abstract}

Clusters of galaxies are the most impressive gravitationally-bound
systems in the Universe and its abundance (the cluster mass function)
is one important statistics to probe the
matter density parameter ($\Omega_m$) and the amplitude of density
fluctuations ($\sigma_8$). The cluster mass function  is usually
described in terms of the Press-Schecther (PS) formalism where the
primordial density fluctuations are assumed to be a Gaussian random
field. In previous works we have proposed a non-Gaussian analytical
extension of the PS approach with basis  on the $q$-power law
distribution (PL) of the non-extensive kinetic theory. In this
paper, by applying the PL distribution to fit the observational mass
function data from X-ray highest flux-limited sample (HIFLUGCS) we
find a strong degeneracy among the cosmic parameters, $\sigma_8$,
$\Omega_m$, and the $q$ parameter from the PL distribution. A joint
analysis involving recent observations from baryon acoustic
oscillation (BAO) peak and Cosmic Microwave Background (CMB) shift
parameter is carried out in order to break these degeneracy and
better constrain the physically  relevant parameters. The present results
suggest that the next generation of cluster surveys will be able to
probe the quantities of cosmological interest ($\sigma_8, \Omega_m$) and the underlying
cluster physics quantified by the $q$-parameter.
\end{abstract}

\pacs{98.80.Es; 95.35.+d; 98.62.Sb}


\maketitle

\section{Introduction}

The recent astronomical observations are strongly suggesting that
the expansion of the Universe is speeding up and not slowing thereby
revealing the presence of some form of repulsive gravity.  The basic set of
experiments includes: observations from SNe Ia\cite{Riess}, cosmic microwave background (CMB) temperature anisotropies\cite{Sper07},
large scale structure\cite{LSS,Eisenstein05}, X-ray data from galaxy clusters\cite{Allen}, age
estimates of globular clusters and old high redshift galaxies\cite{Ages}.
In the present cosmic concordance $\Lambda$CDM model the Universe
is formed of $\sim$ 26\% matter (baryonic + dark matter) and $\sim$ 74\%
of a smooth vacuum energy component. The thermal CMB component
contributes only about 0.01\%, however, its angular power spectrum of
temperature anisotropies encode important information about the structure formation process and other cosmic observables.

On the other hand, the number density of collapsed objects for a given mass at a
certain time, named the \emph{mass or multiplicity function}, is a
key quantity in the analysis of cosmic structures such as clusters
of galaxies. The simplest successful approach to analytically
describe this quantity was developed more than three decades ago by
Press and Schechter\cite{PS74} (hereafter PS). This Gaussian PS
formalism is extensively  adopted to derive the mass function,
$F_{(M)}$, of bounded objects in the observed
Universe\cite{katz94,longair,naga01,reiprich02,reip06,rines08}.

The PS formalism was adopted by Reiprich and
Boringer\cite{reiprich02} in their construction of the cluster mass
function based on the X-ray flux-limited sample of galaxy clusters
(HIFLUGCS) selected from  the ROSAT All-Sky Survey. As a result, the
best fit parameters, $\Omega_{m}=0.12$ and $\sigma_{8}=0.96$, were
obtained in their analysis. These values are, respectively, very low
and very high when compared with the nowadays independent CMB
results\cite{Sper07} thereby leading to some skepticism about the
usefulness of clusters as sensitive cosmological probes. However,
several authors have recently claimed that there is no tension
between cosmological constraints from CMB and
clusters\cite{reip06,rines08}. In particular, Rines and
collaborators\cite{rines08} argued that the dynamical determination
of cluster masses were overestimated and even modest values of
velocity segregation between galaxies and dark matter are sufficient
to match the mass function with the WMAP results.

In this article we discuss a different possibility. We advocate here
that such a discrepancy comes out because the matter density fluctuations
should be described by an intrinsically  non-Gaussian random field (due to the action of
the long-range gravitational interaction)
and, therefore, the PS approach should be somewhat modified.

In previous papers\cite{LM04,ML07}, inspired by the Tsallis
$q$-nonextensive statistics\cite{Tsal88} and kinetic
theory\cite{Plas,Lim2,RS00,ZER}, we have proposed a simple extension
of the PS analytical formalism. Instead the Gaussian function of the
original PS approach, it was adopted the power law (PL) Tsallis
distribution for describing the fluctuations of the density field. A
basic attractive feature of the PL distribution is that the
resulting model is analytically tractable and the standard
Press-Schechter formalism is recovered as particular case. In this
way, a detailed comparison between the two approaches is immediate.

As we shall see, by applying the PL distribution to fit the
observational mass function data from X-ray highest flux-limited
sample (HIFLUGCS) we find a strong degeneracy among the pair of
cosmic parameters ($\sigma_8$, $\Omega_m$), and the $q$ parameter
from PL distribution. Trough a joint analysis involving recent
observations from baryon acoustic oscillation (BAO)
peak\cite{Eisenstein05} and Cosmic Microwave Background (CMB) shift
parameter\cite{Davis,Elgaroy,sze1} the degeneracy is broken and the
tension between independent determinations is alliviated.

\section{The Press-Schechter Approach and the q-Power Law}
\label{Sec1}

\hspace{0.5cm} It is widely known that in the PS original approach
the primordial density fluctuations
$\delta\equiv{\delta\rho}/{\rho}$ for a given mass $M$ is described
by a random Gaussian field
\begin{equation}
P(\delta )=\frac{1}{\sqrt{2\pi }\sigma _{(M)}}\exp \left( -\frac{\delta ^{2}%
}{2\sigma _{(M)}^{2}}\right),  \label{gauss}
\end{equation}
where $\sigma _{(M)}^{2}\equiv \left\langle \delta
_{M}^{2}\right\rangle $ is the mean squared fluctuation. When the
amplitude of the density contrast grows above a critical value
($\delta _{c}$), a bound object is formed  and the fraction
$F_{(M)}$, at a given time, can be written as\cite{PS74,longair}

\begin{equation}
F_{(M)}=\int_{\delta _{c}}^{\infty }P(\delta )\,d\delta
=\frac{1}{\sqrt{2\pi
}\sigma _{(M)}}\int_{\delta _{c}}^{\infty }\exp \left( -\frac{\delta ^{2}}{%
2\sigma _{(M)}^{2}}\right) \,d\delta,  \label{Fm}
\end{equation}
while the distribution of bound objects with masses between $M$ and
$M+dM$ reads

\begin{equation}
\frac{dF_{(M)}}{dM}=+\frac{1}{\sqrt{2\pi }}\frac{\delta _{c}}{\sigma
_{(M)}^{2}}\left( \frac{\partial \sigma _{(M)}}{\partial M}\right)
\exp \left( -\frac{\delta _{c}^{2}}{2\sigma _{(M)}^{2}}\right).
\label{fm3}
\end{equation}

\hspace{0.5cm}Now, if instead of Gaussian initial fluctuations, we
consider that the amplitudes are described by a class of
q-parameterized power law distributions, we have the follow
expression\cite{LM04,ML07,Tsal88,Plas,Lim2}

\begin{equation}
P(\delta )_{PL}=\frac{B_{q}}{\sqrt{2\pi }\sigma _{(M)}}\left[
1-\left( 1-q\right) \cdot \left( \frac{\delta }{\sqrt{2}\sigma
_{(M)}}\right) ^{2}\right] ^{\frac{1}{\left( 1-q\right) }},
\label{prob2}
\end{equation}
where the factor $B_{q}$ is a one-dimensional normalization constant
given by
\begin{enumerate}
\item[a)]  $B_{q}=\left( 1-q\right) ^{\frac{1}{2}}\left( \frac{3-q}{2}%
\right) \frac{\Gamma \left( \frac{1}{2}+\frac{1}{\left( 1-q\right)
}\right) }{\Gamma _{\left( \frac{1}{\left( 1-q\right) }\right) }}$
, (if $0<q\leq 1$)

\item[b)]  $B_{q}=\left( q-1\right) ^{\frac{1}{2}}\frac{\Gamma _{\left(
\frac{1}{\left( q-1\right) }\right) }}
{\Gamma_{\left(\frac{1}{\left( q-1\right) }-\frac{1}{2}\right) }}$
, (if $1\leq q<2$)
\end{enumerate}

In this framework, the non-extensive multiplicity function of bound
objects with masses between $M$ and $M+dM$ reads
\begin{equation}
\frac{dF_{(M)_{PL}}}{dM}=+\frac{B_{q}}{\sqrt{2\pi }}\frac{\delta _{c}}{%
\sigma _{(M)}^{2}}\left( \frac{\partial \sigma _{(M)}}{\partial
M}\right)
\cdot \left[ 1-\left( 1-q\right) \cdot \left( \frac{\delta _{c}}{\sqrt{2}%
\sigma _{(M)}}\right) ^{2}\right] ^{\frac{1}{\left( 1-q\right)}}.
\label{Fm-PL}
\end{equation}
Note that in the limit $q\rightarrow 1$ the above PL expressions
reduce to the ones of the standard Gaussian approach.
\begin{figure}
\includegraphics[width=4.0truein, height=5.0truein,
angle=90]{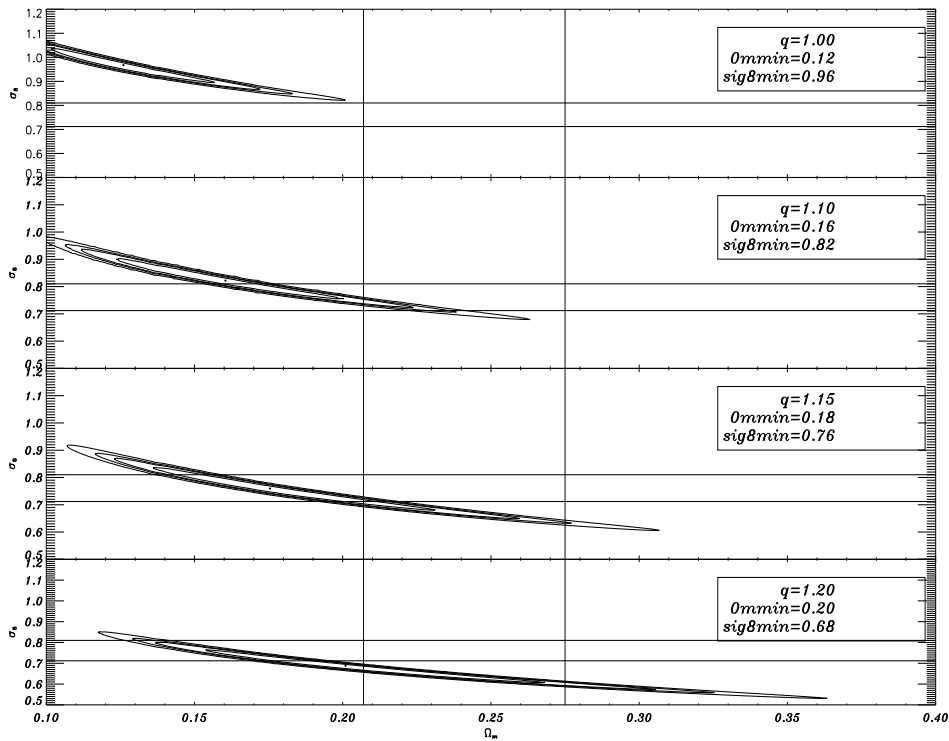} \caption{The First panel shows the results
from the Gaussian PS approach. The results based on the PL
distribution appear in the panels below ($q=1.10$, $q=1.15$, and
$q=1.20$). The solid vertical and horizontal lines show the minimum
and maximum WMAP limits for $\Omega_{m}$ and $\sigma_{8}$,
respectively. Note that the contours using the Gaussian distribution
(the first panel) does not intercept the WMAP values while the ones
based on the PL distribution  are intercepting them for a wide range
of $q$ values ($1<q<1.2$).} \label{Fig1}
\end{figure}

\begin{figure}
\includegraphics[width=2.2truein, height=2.35truein,
angle=90]{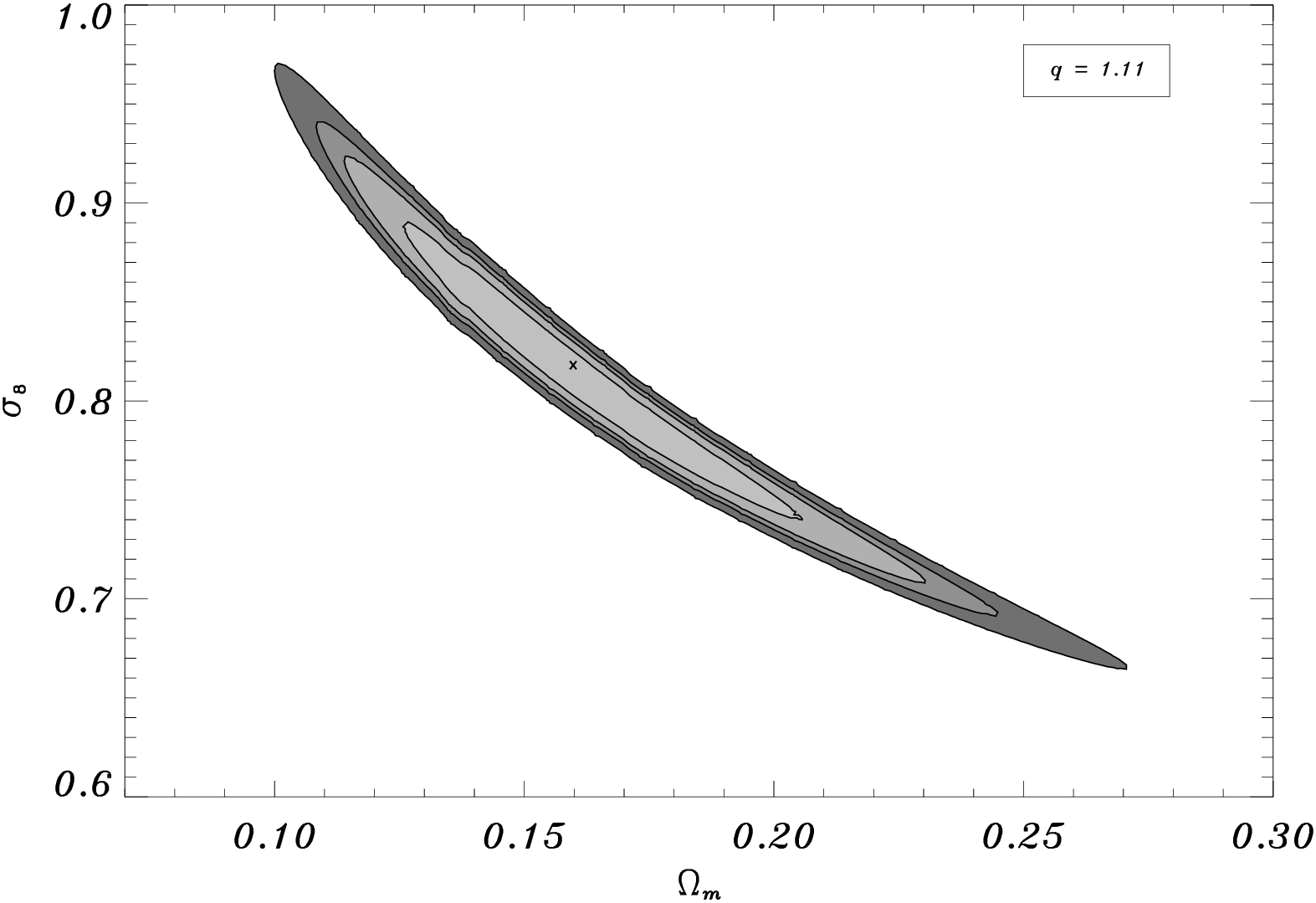}\,\,\,\,\,\,\,\,\,
\includegraphics[width=2.2truein, height=2.35truein, angle=90]{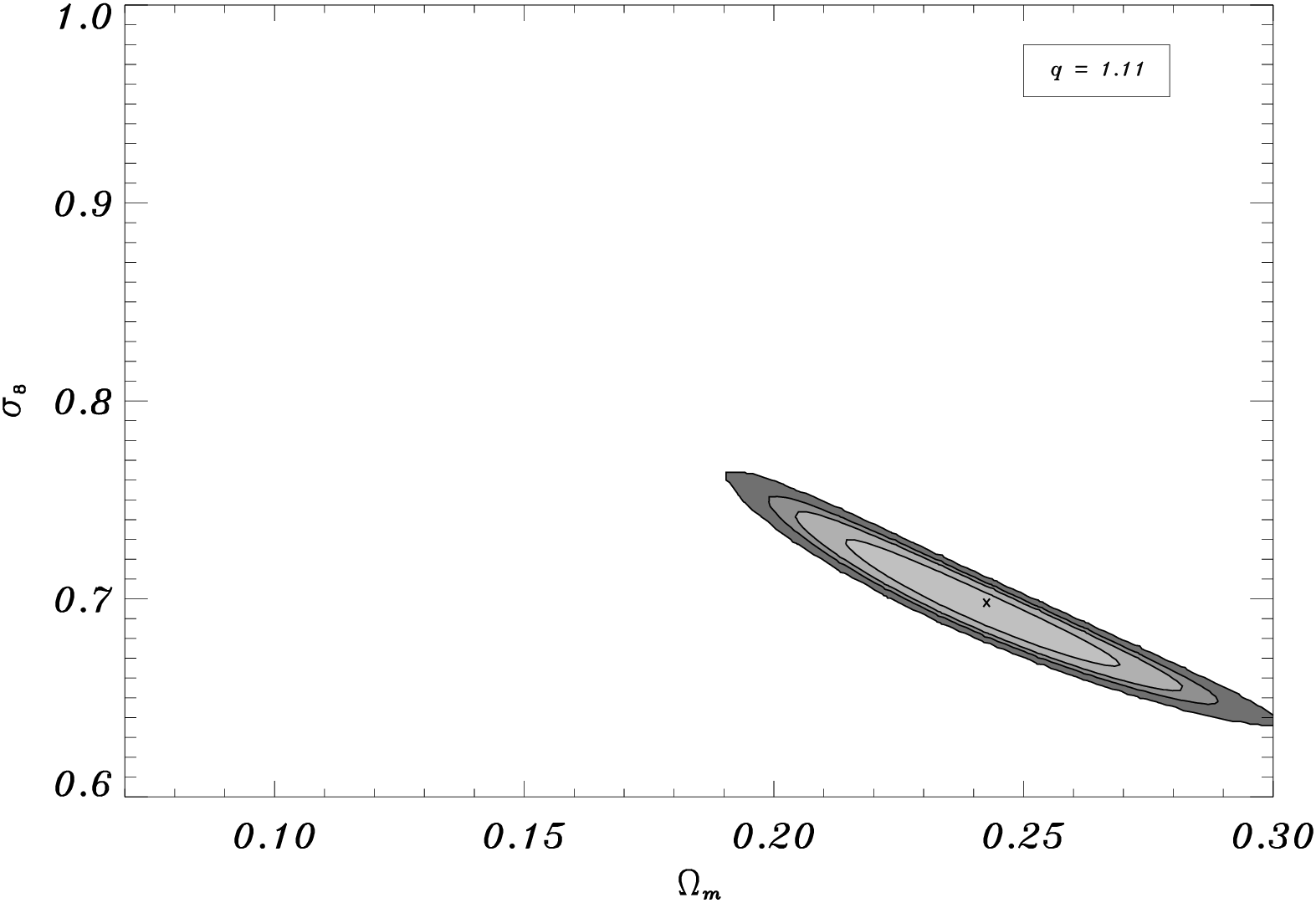}
\caption{Contours on the $\Omega_{m}$-$\sigma_{8}$ plane with
$q=1.11$. The left panel shows poor constraints on the cosmological
parameters. In the right panel we display the results of the joint
analysis with  BAO and Shift Parameter signatures.} \label{Fig2}
\end{figure}

\section{Power Law Distribution and Cosmic Parameters}

As remarked earlier, by performing a $\chi^{2}$ statistical
procedure with  basis on the X-ray HIFLUGCS data sample, Reiprich
and Boehringer (RB)\cite{reiprich02} determined the statistical
confidence contours for the pair of parameters, $\sigma_{8}$ and
$\Omega_{m}$.

In Figure 1 we show the  contours in the $\Omega_{m}$-$\sigma_{8}$
plane obtained by using the standard PS and PL approaches. The first
panel is the RB result. The legend of this panel show the parameter
$q=1.0$, and,  as already explained, when the $q$ parameter from the
PL distribution tends to the unity the Gaussian results are
recovered. In this case, the best-fits of the PS method are also
shown, namely, $\Omega_{m}=0.12$ and $\sigma_{8}=0.96$. The results
from PL distributions ($q\neq1$) can be seen in the panels below
($q=1.10$, $q=1.15$, and $q=1.20$, from top to bottom). Note that
for each panel of Fig.\ref{Fig1}, the solid vertical and horizontal
lines shows the minimum and maximum independent WMAP limits for
$\Omega_{m}$ and $\sigma_{8}$, respectively\cite{Sper07}. The
important point here is that the contours using the Gaussian
distribution (first panel) does not intercept the independent
best-fit values from WMAP even at $99\%$ confidence level. In the
other panels, we see that the $q$ free parameter of the PL
distribution, however, permit the contours to intercept the WMAP
results in a wide range of $q$ values. Roughly estimates show that
the range between $q=1.06$ and $q=1.2$ is allowed, and, in
principle, such a degeneracy need to be removed.

\section{Joint Analysis and Discussion}\label{Sec2}

In principle, the above results suggest that in the non-extensive framework
there are many possibilities for the theoretical mass function.
Still more important, some of them are working in the right direction and may help for reconciling
the independent estimates of the
cosmic parameters.

In this section we discuss how  the parametric 2-dimensional spaces
$\sigma_{8}-\Omega_m$, $\Omega_{m}-q$ and $\sigma_{8}-q$  can be further constrained by apllying a
statistical analyses involving different cosmological
observations. To this end we consider the current estimates of the baryon
acoustic oscillations found in the SDSS data \cite{Eisenstein05}, as
well as, the shift parameter from WMAP observations \cite{Sper07}.
The basic aim is to break the degeneracy between the $\Omega_{m}$,
$\sigma_{8}$ and $q$ parameters in order  to better constrain the PL
distribution that fits the HIFLUGCS data.

\subsection{BAO}

The Baryon Acoustic Oscillations (BAO) in the primordial baryon-photon fluid,
leave a characteristic signal on the galaxy correlation function, a bump at a scale
$\sim$  100 Mpc, as observed by Eisenstein and coworkers\cite{Eisenstein05}. This
signature furnishes a standard rule which can be used to constrain
the following quantity:
\begin{eqnarray}
{\cal{A}} \equiv \frac{\Omega_m^{1/2}}{
{{\cal{H}}(z_{\rm{*}})}^{1/3}}\left[\frac{1}{z_{\rm{*}}}
\Gamma(z_*)\right]^{2/3}  = 0.469 \pm 0.017, 
\label{A}
\end{eqnarray}
where ${\cal{H}} = H(z)/H_0$ is the normalized Hubble parameter of
the $\Lambda$CDM model, $z_*=0.35$ is a typical redshift of the SDSS
sample, and $\Gamma(z_*)$ is the dimensionless comoving distance to
the redshift $z_*$.  This quantity can be used even for more general models which
do not present a large contribution of dark energy at early
times\cite{DST07}.

\subsection{CMB shift parameter}

A useful quantity to characterize the position of the CMB power
spectrum first peak is the shift parameter.  For a flat Universe it
is given by\cite{Elgaroy,efstathiou}:
\begin{equation}
 {\cal R}=\sqrt{\Omega_m}\int_0^{z_r}\frac{dz}{{\cal H}(z)} = 1.70 \pm 0.03\;,
\end{equation}
where $z_r = 1089$ is the recombination redshift and the value for
${\cal R}$ above is calculated from the MCMC of the WMAP 3-yr in the
standard flat $\Lambda$CDM model.

\begin{figure}
\includegraphics[width=2.2truein, height=2.35truein,
angle=90]{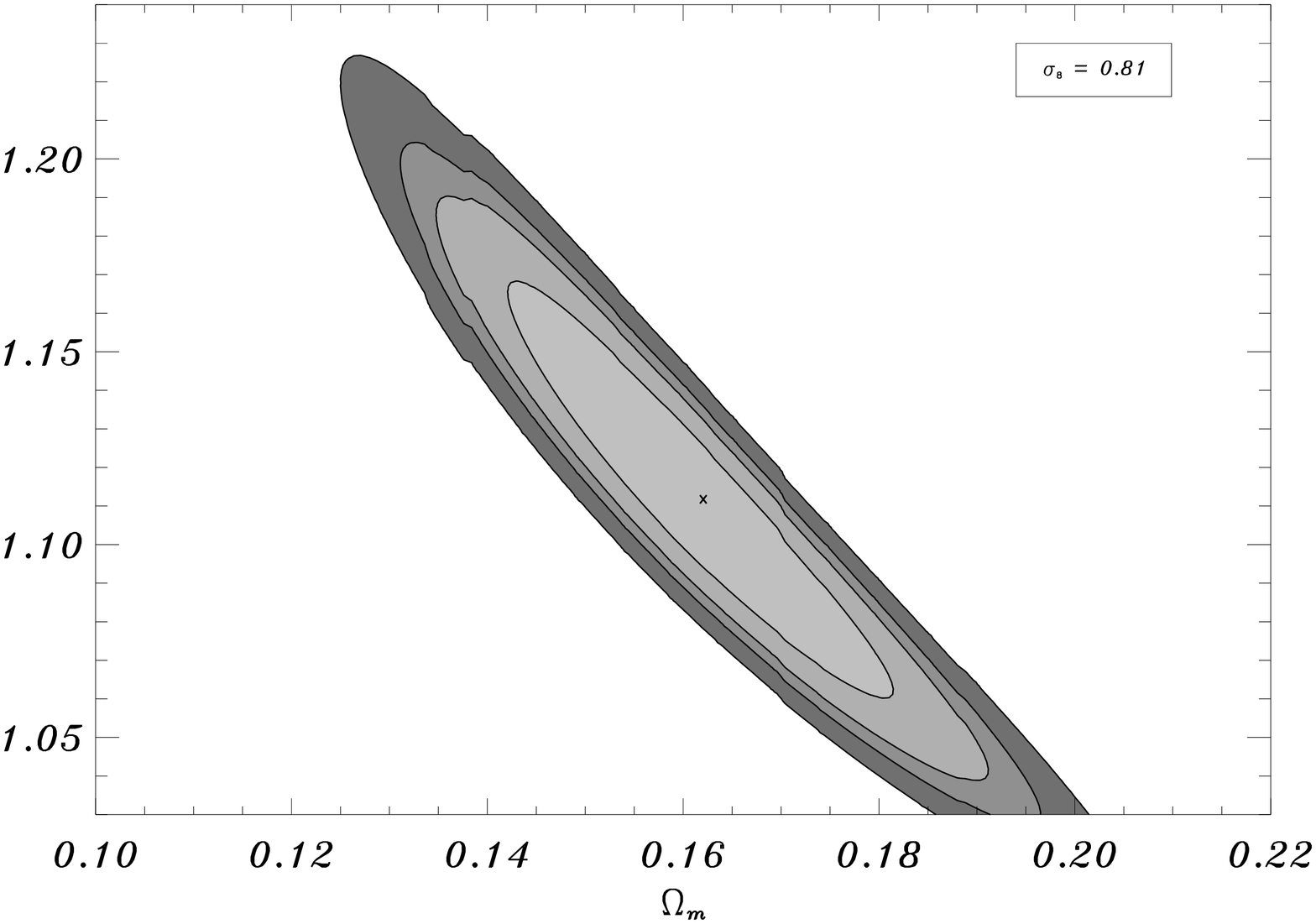}\,\,\,\,\,\,\,\,\,
\includegraphics[width=2.2truein, height=2.35truein, angle=90]{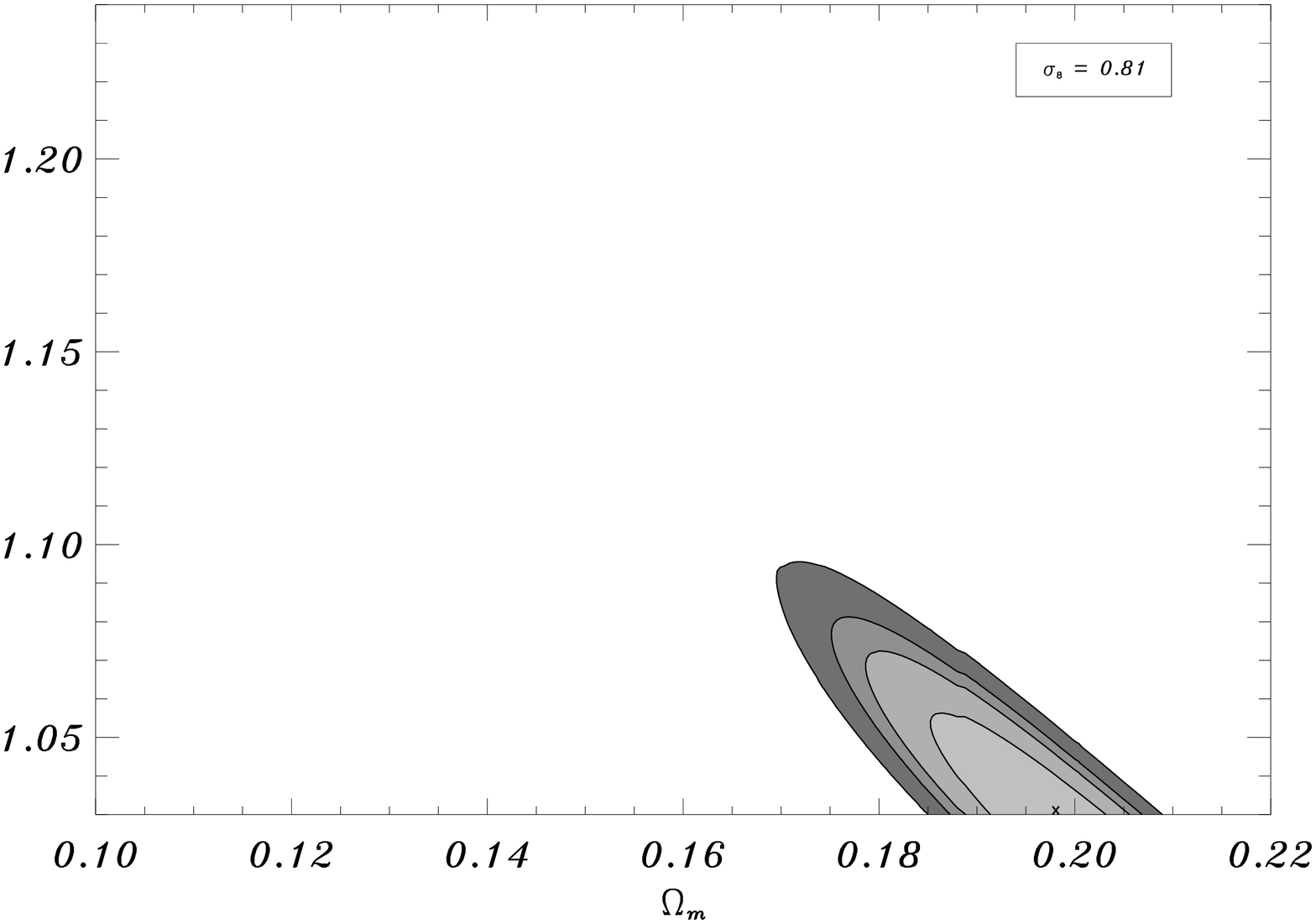}
\caption{Contours on the $\Omega_{m}$-$q$ plane with
$\sigma_{8}=0.81$. In the left panel the contour permits almost
all values for the $q$ parameter. The right panel shows the joint
analysis with the BAO and the Shift Parameter: the $q$ parameter
is restricted in the range $1.0<q<1.10$, and we have a high
constraint on the $\Omega_{m}$ parameter as well.} \label{Fig3}
\end{figure}

\subsection{Results}

Let us know discuss the main results of our statistical analyses. In
Fig. \ref{Fig2}, by fixing $q=1.11$, we show the contours on the
$\Omega_{m}$-$\sigma_{8}$ plane. The left panel shows that we have
poor constraints on these cosmological parameters. However, when a
joint statistical analysis is performed using the BAO the shift
parameter (right panel) the available space parameter is
considerably reduced. Note that the contour becomes small and
shifted to higher and lower values of $\Omega_{m}$ and $\sigma_{8}$,
respectively.

By applying the same procedure (now fixing $\sigma_{8}=0.81$) and
plotting the contours on the $\Omega_{m}$-$q$ plane, the left panel
of Fig. \ref{Fig3} is obtained. One may see that almost all values
for the $q$ parameter is allowed, however, a joint analysis with BAO
and shift parameter restricts severely the  $q$ parameter ($1 \leq q
\leq 1.1$).  Note also that the $\Omega_{m}$ parameter becomes
tightly constrained.

\begin{figure}
\includegraphics[width=2.2truein, height=2.35truein,
angle=90]{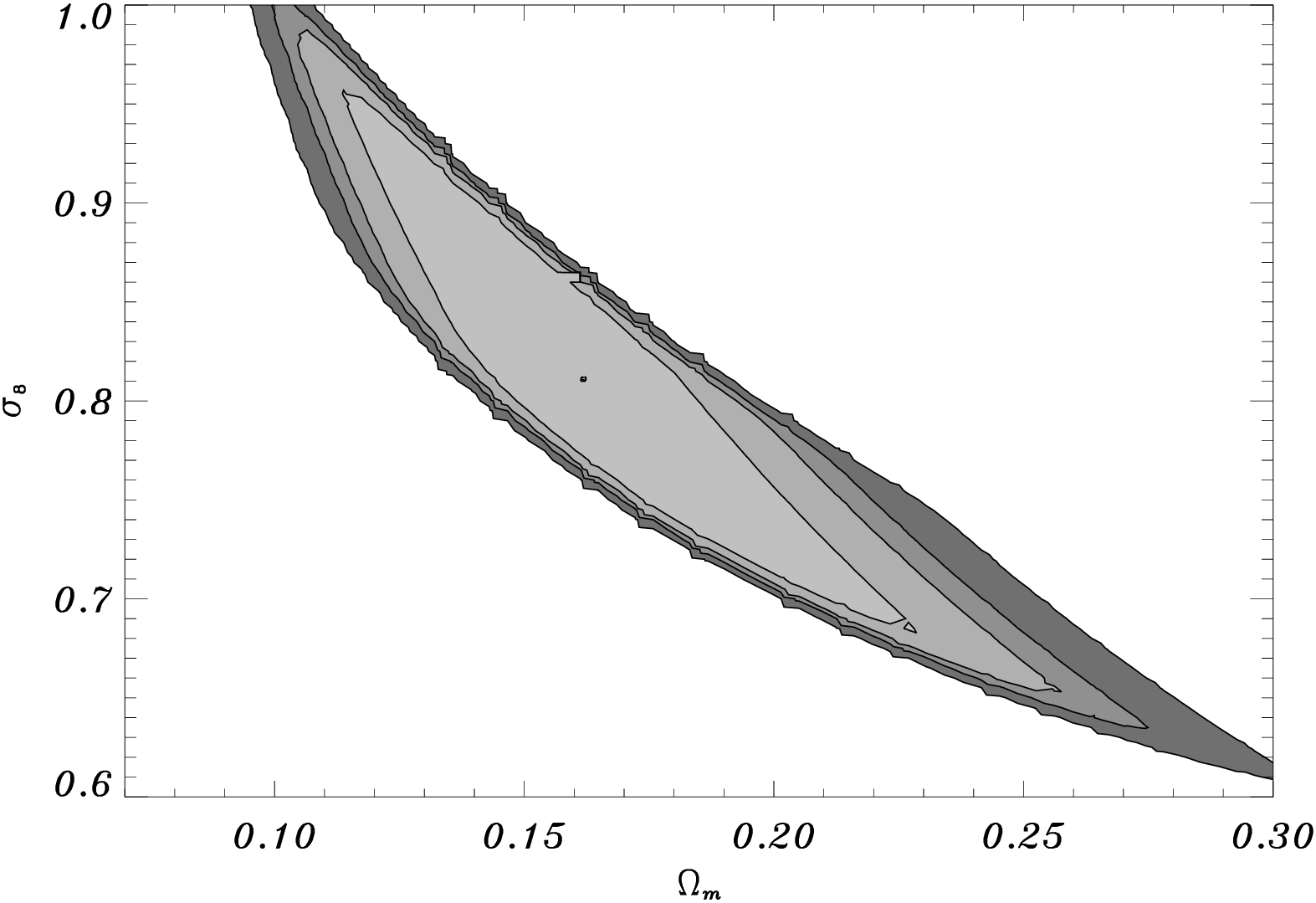}\,\,\,\,\,\,\,\,\,
\includegraphics[width=2.2truein, height=2.35truein, angle=90]{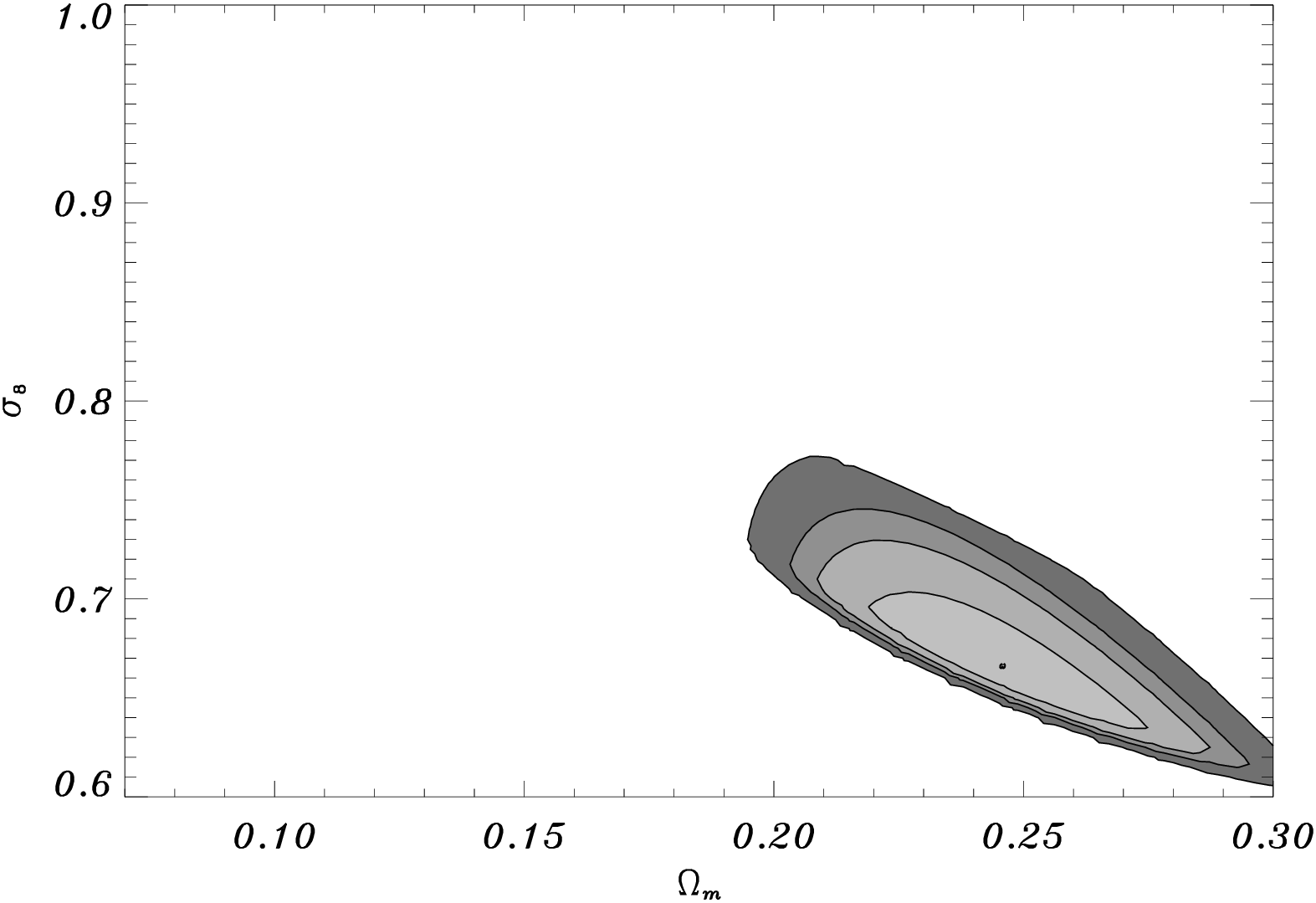}
\caption{The left panel shows the contour on the
$\Omega_{m}$-$\sigma_{8}$ plane when we marginalize over all the allowed
values of $q$. In the right panel we apply the BAO and the Shift
Parameter signatures. The cosmological parameters are in good agreement
with the WMAP limits.} \label{Fig4}
\end{figure}

Finally, we discuss the contours on the $\Omega_{m}$-$\sigma_{8}$
plane when we  marginalize over all possible values of $q$. The
result is shown  on the left panel of Fig. \ref{Fig4}. In the right
panel we display the result of the joint analysis. It is interesting
that by applying BAO and shift parameter signature the space
parameter is extremely reduced and displaced as occurred in Fig.
\ref{Fig2} (see also right panel there). We stress that we  have
marginalized over $q$ and performed a joint analysis. The main
consequence is that the  cosmological parameters are now in
agreement with the latest WMAP results\cite{Sper07}.

\section{Conclusions}
\label{Sec3}

In the precision era of cosmology it is very important to show the
compatibility of the cosmological parameters as determined from
independent observations.  The current standard cosmological model,
i.e., a flat, accelerating Universe composed of $\simeq 1/3$ of
matter (baryonic + dark) and $\simeq 2/3$ of a dark energy component
in the form of the vacuum energy density ($\Lambda$) seems to be
fully consistent with a variety of observational data. However, some
tension has recently been detected between the determinations of the
pair ($\sigma_8,\Omega_m$) from galaxy clusters and CMB. In
principle, many temporary solutions are possible, but, hopefully,
only one will survive to future analyses based on the next
generation of cluster surveys.

In this article we have discussed an alternative route which
seems interesting to be investigated from a theoretical and observational viewpoint.
Initially, inspired by the non-extensive kinetic theory, we have extended the original
Gaussian distribution of the primordial density
field\cite{LM04,ML07}. By using this PL distribution to fit the
observational mass function from the HIFLUGCS data we have
identified a degeneracy involving the cosmic parameters $\sigma_8$
and $\Omega_m$ with the $q$ parameter from the PL distribution.

Finally,  a joint analysis involving the baryon acoustic oscillation
signature and the  CMB  shift parameter have been applied. As a result, we have shown
that the non-extensive PL distribution may alliviate the tension
underlying the independent determinations of the cosmic parameters
(see right panel of Fig. \ref{Fig4}).

\section*{Acknowledgements}
The authors are grateful to S. H. Pereira and J. M. Silva for
helpful discussions. LM and JVC are supported by FAPESP No.
07/00036-4 and 05/02809-5, respectively, and JASL is partially
supported by CNPq and FAPESP grant No. 04/13668 .

\end{document}